# Room-Temperature Electronically-Controlled Ferromagnetism at the LaAlO$_3$/SrTiO$_3$ Interface

Feng Bi,[1] Mengchen Huang,[1] Chung-Wung Bark,[2] Sangwoo Ryu,[2] Chang-Beom Eom,[2] Patrick Irvin[1] and Jeremy Levy[1*]

[1] Dept. of Physics & Astronomy, University of Pittsburgh, Pittsburgh, Pennsylvania, 15260, USA.

[2] Dept. of Materials Science and Engineering, University of Wisconsin-Madison, Madison, Wisconsin, 53706, USA.

*jlevy@pitt.edu

**Reports of emergent conductivity, superconductivity, and magnetism have helped to fuel intense interest in the rich physics and technological potential of complex-oxide interfaces. Here we employ magnetic force microscopy to search for room-temperature magnetism in the well-studied LaAlO$_3$/SrTiO$_3$ system. Using electrical top gating to deplete electrons from the oxide interface, we directly image an in-plane ferromagnetic phase with sharply defined domain walls. Itinerant electrons, introduced by a top gate, align antiferromagnetically with the magnetization, at first screening and then destabilizing it as the conductive state is reached. Subsequent depletion of electrons results in a new, uncorrelated magnetic pattern. This newfound control over emergent magnetism at the interface between two non-magnetic oxides portends a number of important technological applications.**

Vigorous efforts have been made to integrate magnetism with semiconductors[1-3]. Ideally, a ferromagnetic semiconductor would possess full electrical control over its magnetic properties combined with strong coupling to the spin of mobile charge carriers. Efforts to identify suitable materials have focused on diluted magnetic semiconductors (DMS) such as (Ga,Mn)As[4], diluted magnetic oxides (DMOs)[3], and magnetoelectric materials such as chromia[5]. Strong electronic correlations[6] can also induce charge-ordered phases, produce electronic phase separation, and stabilize various types of magnetic order.

The two-dimensional electron liquid (2DEL) that forms at the interface between the two insulating non-magnetic oxides LaAlO$_3$ (LAO) and SrTiO$_3$ (STO)[7] has drawn widespread attention due to its possession of a remarkable variety of emergent behavior including superconductivity[8], strong Rashba-like spin-orbit coupling[9,10], and ferromagnetism[11-15]. The first signatures of magnetism at the LAO/STO interface were reported in magnetotransport measurements by Brinkman *et al*.[11] DC scanning quantum interference device (SQUID) magnetometry measurements by Ariando *et al.*[12] reported ferromagnetic hysteresis extending to room temperature. Torque magnetometry measurements by Li *et al.*[13] showed evidence for in-plane magnetism with a high moment density (~0.3 $\mu_B$/unit cell). Scanning SQUID microscopy by Bert *et al.*[14] revealed inhomogeneous micron-scale magnetic "patches". X-ray circular dichroism measurements by Lee *et al.* indicate that the ferromagnetism is intrinsic and linked to $d_{xy}$ orbitals in the Ti $t_{2g}$ band[15]. Despite this evidence, the existence and nature of magnetism in LAO/STO heterostructures has remained controversial. Neutron reflectometry measurements by Fitzsimmons *et al*. on LAO/STO superlattices[16] found no magnetic signatures; their measurements established a bulk upper limit thirty times lower than what was reported by Li *et al.*[13] Salman *et al.* reported relatively small moments from LAO/STO superlattices (~$2\times10^{-3}$ $\mu_B$ /unit cell) using β-detected nuclear magnetic resonance[17].

Our search for magnetism at the LAO/STO interface is guided by the fact that most of the interesting behavior observed at the LAO/STO interface—superconductivity, spin-orbit coupling, anisotropic magnetoresistance, and anomalous Hall behavior—depends strongly on carrier density[18,19,20,21]. The electron density at the interface can be controlled using a number of techniques including back-gating[22], top-gating[10], polar adsorbates[23], or via nanoscale control using conductive atomic force microscopy (AFM) lithography[24].

Here we investigate magnetism at the LAO/STO interface using magnetic force microscopy (MFM)[25]. By using top-gated LAO/STO heterostructures we are able to search for magnetism as a function of interfacial carrier density. All measurements are performed in ambient conditions at room temperature. The sample is kept within a darkened chamber during experiments in order in minimize photoexcitation of carriers. LAO/STO heterostructures are fabricated by depositing 12 unit cell (u.c.) LAO films on TiO$_2$-terminated [001] STO substrates using pulsed laser deposition with *in situ* high-pressure reflection high energy electron diffraction (RHEED). LAO/STO heterostructures are patterned with top circular electrodes and concentric arc-shaped interfacial contacts, sketched in Fig. 1a. Details of the growth techniques and device fabrication are described more fully in the Supplementary Information (SI). Experimental measurements are presented for three devices whose parameters are summarized in Table S1. For consistency, the main results are presented for Device A, while selected similar results for the other two devices are included in the SI.

A two-terminal capacitor device is used to electrically gate the LAO/STO interface (Fig. 1a). The top circular electrode is grounded and a voltage $-V_{dc}$ is applied to the annular interface contact. (This configuration is equivalent to grounding the interface and applying $+V_{dc}$ to the top electrode.) Decreasing $V_{dc}$ depletes the interface of mobile electrons, while increasing $V_{dc}$ leads to electron accumulation and results in a conductive interface. The critical voltage for the metal-insulator transition (MIT) is device-dependent and generally exhibits voltage hysteresis in the range 0 V to -2 V. The transition is readily identified as an inflection point in the capacitance-voltage (CV) spectrum (Fig. S2).

A CoCr-coated AFM tip is magnetized using electromagnets with magnetic field up to 2000 Oe. The cantilever has resonant frequency $f_0$ and free amplitude $A_0$; in MFM mode it is mechanically driven by a piezoelectric transducer near $f_0$. When the tip is placed in proximity to a sample, the cantilever's resonant motion is altered in ways that can be traced directly to the force gradient $\partial F_z / \partial z$, which in turn produces changes in the amplitude $\Delta A$, phase $\Delta \phi$ and frequency $\Delta f$ of the cantilever resonance[25,26]:

$$\Delta A \approx \frac{2A_0 Q}{3\sqrt{3}k} \cdot \frac{\partial F_z}{\partial z} \quad (1a) \qquad \Delta \phi \approx \frac{Q}{k} \cdot \frac{\partial F_z}{\partial z} \quad (1b) \qquad \Delta f_0 \approx -\frac{1}{2k} \cdot \frac{\partial F_z}{\partial z} \cdot f_0 \quad (1c)$$

In Eq. 1(a-c), $k$ is the spring constant of the cantilever and $Q$ is the quality factor of the resonance. Force gradients may arise from magnetic or non-magnetic interactions, so it is important to conduct experiments that can distinguish the two sources of contrast.

MFM imaging using frequency modulation[26] is performed directly over the top electrode (Fig. 1a). The MFM tip is electrically grounded and shielded by the grounded top gate, eliminating any possible electrostatic coupling between the tip and electrode. The topographic image in Fig. 1b has nanometer-scale surface roughness associated with inhomogeneity in the Au layer. The tip is magnetized perpendicular to the LAO/STO interface, and cantilever frequency-shift images are obtained for conducting ($V_{dc}$ = 0 V, Fig. 1c) and insulating ($V_{dc}$ = -2 V, Fig. 1d) states of the LAO/STO interface. The frequency shift is negligibly small for both voltage gating conditions, with the exception of a few topographic features. The RMS frequency shift is $\Delta f$=0.43 Hz for $V_{dc}$=0 V and $\Delta f$=0.44 Hz for $V_{dc}$=-2 V, close to the noise floor of 0.27 Hz for the measurement. The tip is then magnetized along the in-plane [010] orientation, and frequency-shift images are acquired with the same tip, over the same region on the sample, under the same voltage bias conditions as with Fig. 1c-d. In the conductive state ($V_{dc}$ = 0 V, Fig. 1e), the image shows negligible spatial variations in the frequency shift. In the insulating state ($V_{dc}$ = -2 V, Fig. 1f), the frequency-shift image shows significant contrast, of order 20 Hz peak-to-peak (Note the change in frequency-shift scale.). The boundaries between domains of approximately equal frequency shift are sharp, straight, and aligned approximately 30-40 degrees relative to the horizontal axis. Experiments performed at various angles between the MFM tip and the sample show that the stripe contrast is maximized for tip magnetization oriented along the [100] or [010] direction, and vanishes for tip magnetizations that approach the $[110]$ and $[\bar{1}10]$ direction (Fig. S9). Experiments performed with a non-magnetic tip show no visible contrast at any gate voltage in the range -3 V to +3 V (Fig. S18).

To rule out possible influences from metal deposition and possible local leakage currents from the top electrode, MFM experiments are also performed directly over the LAO surface in a region ~10 μm from the edge of the circular top electrode (Fig. 2a). The exposed LAO surface is atomically flat with clearly-resolved single-unit-cell 0.4 nm steps. To check that electrical gating is effective even tens of micrometers away from the circular electrode edge, the surface potential is mapped using Kelvin probe force microscopy (KFM). Fig. 2b shows the surface potential distribution over the investigated area with top gate grounded and the interface biased at 3 V. (The work function difference between the metal top gate and the LAO surface is accounted for, and described in Fig. S8.) A cross-sectional profile analysis shows that the electrode-induced top gating extends far from the top electrode (Fig. 2c). Within the rectangular area marked by a dashed line, MFM measurements are performed as a function of gate voltage (Fig. 2d). The first MFM image, taken at $V_{dc}$ = -4 V ("State 1"), shows strong contrast in the frequency channel similar to that seen over the electrode. No correlation of the frequency shift with unit-cell terraces is observed. As the gate bias is increased, the contrast between domains diminishes. For $V_{dc}$ = -2 V the domain contrast has nearly vanished, with new horizontal bands appearing parallel to the fast scan axis. For $V_{dc}$ = 0, the contrast is absent. Subsequent decrease of $V_{dc}$ restores the magnitude of the response but with a new domain pattern that is uncorrelated with the previous one ("State 2"). The slight decrease in magnitude is correlated with charge hysteresis in the CV spectrum (Fig. S2) during voltage gate sweeps.

The experimental results presented thus far provide compelling evidence for electronically controlled ferromagnetism. The sensitivity to the electronic state of the interface, and the sharpness of the magnetic features, indicate that the magnetism resides at the LAO/STO interface. In the experiments described below, the patterns will be interpreted as ferromagnetic domains and their properties will be further explored.

The observed large-scale domain structures in the MFM images are generally reproducible from one scan to the next (Fig. 3a-b), but there are differences in fine domain structures from successive scans that suggest tip-induced magnetization switching. The subtle changes from images acquired in succession are more readily seen by examining the phase error channel $\Delta\phi$ (Fig. 3c-d) from the same data set. A line cut perpendicular to the domain boundaries (Fig. 3e) shows that the first image taken at $V_{dc}$=-2 V exhibits domains that are 40-50 nm in width; this fine domain structure is largely absent in the subsequently acquired image.

The gate dependence of the MFM images demonstrates a strong interaction between ferromagnetism and itinerant electrons. To help explore this connection, we perform dynamic magneto-electric force microscopy (MeFM) experiments (Fig. 4a). Unlike MFM, the tip is not mechanically driven; instead, the gate is electrically modulated at the mechanical resonance of the cantilever, and the resulting ac magnetic field from the LAO induces resonant motion of the magnetized cantilever. The surface topography image is included in Fig. 4b. MeFM amplitude images (Fig. 4c) are acquired at values of $V_{dc}$ ranging between -3.5 V and 0 V. As the carrier density increases, the contrast in MeFM images becomes weaker and more diffuse. Similar contrast is also observed in MeFM images taken over the top electrode (Fig. S11-12) and MeFM phase images (Fig. S14).

Two-dimensional Fourier analysis of the MeFM images (Fig. S15) shows that the domain wall width diverges as the conductive phase is reached. The observed stripe domains and the contrast changes in MeFM images agree qualitatively with the results obtained by conventional MFM imaging. The coupling of the magnetic response to carrier density establishes that the electrons entering the interface become spin-polarized, aligning antiferromagnetically with the magnetic domains. Additional scan sequences (Fig. S13-14) show that voltage hysteresis in the magnetic response can be traced back to observed hysteresis in the CV spectrum (Fig. S2).

The MFM tip strongly perturbs the magnetic domain structure, as can be seen from successive scans (e.g., Fig. 3). Angle-dependent MFM measurements (Fig. S9) can be interpreted as evidence for in-plane anisotropy along the $[100]$ and $[010]$ directions, but it is more likely that the domains are aligned along the $[110]$ and $[\bar{1}10]$ directions, with domain structure that is stable against perturbation from the MFM tip only close to $\theta=0°$. The domain walls themselves are generally sharp and resolution-limited, with widths that depend on $V_{dc}$ (Fig. S15). The absence of features in MFM images with vertically magnetized tips (Fig. 1c-d) have at least two possible explanations. One is that the domain walls are Néel-type, with magnetization rotating in the plane of the sample or vanishing at the domain wall. An alternate explanation is that the magnetization of the tip is too strongly perturbing of the domain state to allow them to be imaged.

Theories of magnetism at the LAO/STO interface generally invoke localized unpaired $d_{xy}$ electron spins at the interface that couple via exchange with itinerant carriers. The mechanism depends on the relative density of localized versus itinerant electrons, as well as their orbital character[20,27,28]. Fidkowski et al. describe a ferromagnetic Kondo model[28] in which local $d_{xy}$ moments couple ferromagnetically to delocalized $d_{xz}/d_{yz}$ carriers; models with ferromagnetic exchange are also described by Joshua et al.[20] and Bannerjee et al.[29]. Michaeli et al.[30] describe a model based on Zener (antiferromagnetic) exchange between localized and delocalized $d_{xy}$ carriers[31].

The local moments themselves are postulated to arise either from electronic correlations[32] or interfacial disorder[28,30], or are related to oxygen vacancies[27,33]. Extrinsic sources of magnetic impurities have been ruled out experimentally[11,12,15]. The decrease in net magnetization with increasing carrier density (Fig. S16c) indicates that the exchange is primarily antiferromagnetic and distinct from the Kondo-related[34] ferromagnetism[35] reported for electrolyte-gated STO. The source of the local moments is not obviously constrained by these experiments, except for the fact that they appear to be highly uniform-domain walls are sharp and highly linear and not pinned by fluctuations in moment density. The two growth conditions explored here (see Table S1) are both believed to have a low density of oxygen vacancies.

The existence of high-temperature ferromagnetism at the insulating LAO/STO interface offers a way to resolve many of the contradictory reports regarding magnetism at the LAO/STO interface. Most experimental investigations have been performed with conducting interfaces, a regime for which high-temperature ferromagnetism is suppressed. Any inhomogeneity that locally depletes the interface, e.g. defects or surface adsorbates[23], could give rise to local insulating regions that exhibit magnetic behavior. Similarly, LAO/STO structures grown at pressures close to the insulating transition (e.g., P(O$_2$)=10$^{-2}$ mbar for Ref.12) may contain local insulating regions that exhibit room-temperature ferromagnetic behavior.

Based on these results, we can postulate a phase diagram for the types of magnetism observed at the LAO/STO interface. The phase diagram does not purport to quantify the precise ferromagnetic transition temperature $T_c$ or carrier density dependence for each phase; rather, is helpful for discussing possible physical explanations of the observed results and placing them in context with other reports in the literature. Fig. 5a-d illustrates four distinct postulated phases with varying types of charge, spin, and orbital order as a function of temperature and total carrier density.   In region A, the carrier density at the interface is too low to support magnetism.  For example, the LAO layer might be below the critical thickness for magnetism (as observed by Kalitsky *et al.*[36]). In region B, the room-temperature ferromagnetic phase exists. Local $d_{xy}$ moments order via antiferromagnetic exchange with "semi-localized" $d_{xy}$ electrons. In region C, the conducting phase, the $d_{xy}$ electrons are fully extended, and ferromagnetism is suppressed.  In region D, new types of carriers with $d_{xz}$/$d_{yz}$ are introduced at the Lifshitz transition[19], resulting in a second magnetic phase M*, which is associated with superconductivity, strong spin-orbit coupling, anisotropic magnetoresistance, and anomalous Hall effects.  The moments are predicted to align ferromagnetically with the $d_{xz}$/$d_{yz}$ electrons, from Hund's rule coupling[20,28,29]. There are quantum phase transitions at three critical densities: $n_0$, where FM order begins at low temperature, $n_{MIT}$, the metal-insulator transition, and $n_L$, the Lifshitz transition.  The most important point to emphasize is that there are two distinct magnetic phases, and thus two classes of theories that are required to understand the rich magnetism observed in this system.

There are many unresolved questions regarding the nature of the FM state.  First and foremost, the density of localized moments, and the relative density of delocalized carriers, is not well characterized by these MFM measurements.  The magnetic moment density is a quantity that in principle can be obtained from MFM measurements but is challenging here given the fact that the magnetization is so strongly perturbed by the

MFM tip. The magnetic easy axes are not readily identified, although there is clear in-plane magnetic anisotropy. Future refinements of these experiments as well as new ones will undoubtedly answer these questions and help to constrain theoretical descriptions.

The discovery of electrically-controlled ferromagnetism at the LAO/STO interface at room temperature provides a new and surprising route to a wide range of spintronics applications. Many effects—not yet demonstrated—are nevertheless expected, such as spin-torque transfer, spin-polarized transport, electrically controlled spin-wave propagation and detection, magnetoresistance effects, and spin-transistor behavior. This versatile spintronic functionality may also be combined with conductive AFM control over the metal-insulator transition[24], both for room temperature spintronics applications and low-temperature quantum devices.

**METHODS SUMMARY**

Samples are grown by pulsed laser deposition. Before deposition, low-miscut (<0.1°) STO substrates are etched using buffered HF acid to keep the $TiO_2$-termination. Then the STO substrates are annealed at 1000 °C for several hours so the atomically flat surfaces are created. During the deposition, a KrF exciter laser ($\lambda = 248$ nm) beam is focused on a stoichiometric LAO single crystal target with energy density 1.5 J/cm$^2$ and each LAO unit cell is deposited by 50 laser pulses. Two different growth conditions are used for the substrate growth temperate $T$ and chamber background partial oxygen pressure $P(O_2)$: (1) $T$=550 °C and $P(O_2)$=10$^{-3}$ mbar ; (2) $T$=780 °C, $P(O_2)$=10$^{-5}$ mbar. For samples grown in condition (2), after deposition they are annealed at 600 °C in $P(O_2)$=300 mbar for one hour to minimize oxygen vacancies. The device fabricated is a two-terminal capacitor structure, consisting of a circular top electrode and an arc-shaped electrode contacting the LAO/STO interface, depicted schematically in Fig. 1a.

Conventional dynamic MFM is performed using either slope-detection method[25] or frequency-modulation method[26]. In the main text, frequency modulation is employed. A 90° phase shift between cantilever drive and response is maintained using a feedback loop. The process variable $\Delta\phi$, the difference between the measured phase shift and the target value of 90°, is sensitive to rapid changes in magnetization that are similar to the topographic error signal in conventional AFM topographic imaging. This error signal is displayed in Fig. 3. The mean tip-surface separation over the top electrode is 30 nm during the MFM imaging. For the MFM experiments above the exposed LAO surface (Fig. 2), the tip-surface separation is 20 nm and the scan size is 3 µm. In both Fig. 1 and Fig. 2, the cantilever's drive frequency is $\omega_0/2\pi$=70 kHz. The amplitude, phase, and frequency shift are all recorded for various values of $V_{dc}$.

Dynamic magneto-electric force microscopy (MeFM) is also employed in our experiments. Rather than driving the tip mechanically, the sample is driven with a static and sinusoidally modulated voltage. The gate-modulated sample magnetization periodically changes the magnetic force, driving the tip resonantly and producing a cantilever oscillation that is detected with a lock-in amplifier. In Fig. 4, the tip remains magnetized along the [010] direction. The top electrode is grounded to minimize electrostatic interactions between the tip and sample, and a combined voltage is applied to the interface $V_{IF}(t) = -(V_{dc} + V_{ac}\sin(2\pi ft))$, with $V_{ac}$=0.14 V and $f$=70.6 kHz. The tip is maintained at a distance of 20 nm above the LAO surface and scanned over a 1.5 µm square area on Device A. The deflection signal from the tip is monitored by a lock-in amplifier at the resonant frequency $f$.

Figure Legends:

**Fig. 1 | MFM experiments on gated LaAlO$_3$/SrTiO$_3$ heterostructures. a,** Sketch of experimental setup. The MFM tip is mechanically driven by a piezoelectric transducer near its resonant frequency and kept at a constant height Δh above the surface. The top electrode is grounded and a dc bias is directly applied to the interface. During the experiment, frequency modulation is used to lock the phase around 90°. The tip resonant frequency is tracked; the frequency shift is proportional to the magnetic force gradient. **b,** Actual surface morphology profile over the investigated 1.4 μm×1.4 μm area on top electrode. **c,d,** MFM frequency images using vertically magnetized tip with V$_{dc}$ =0 V and -2 V. **e,f** MFM frequency images over the same area using tip that magnetized horizontally along the [010] direction. Images are taken under V$_{dc}$ =0 V and -2 V.

**Fig. 2 | MFM experiments performed over the exposed LAO region. a,** experimental setup, similar to that of Fig. 1a, except the scanning location is now over the exposed LAO approximately 10 μm away from the electrode edge. **b,** Kelvin probe force microscopy measurement of a region that includes the area for which MFM measurements are made. The voltage bias is V$_{dc}$=3 V. The topography is shown as height, while the color maps onto the measured surface potential (the work function is already subtracted). **c**, a linecut of the surface potential map, showing that voltage gating decays but is reduced by ~1 V in the area over which MFM images are acquired. **d,** MFM frequency images over a 3 μm×3 μm area indicated by the black dashed line enclosed region in **b**. The MFM tip is magnetized horizontally parallel to the [010] sample direction. MFM frequency images for V$_{dc}$ increasing from -4 V to 0 V then decreasing back to -4 V. Magnetic domain features are clearly observed for V$_{dc}$<-2 V. The observed domain patterns and contrast changes with respect to V$_{dc}$ qualitatively agree with the results in **Fig. 1f**.

**Fig. 3 | MFM images acquired in succession. a,b,** two successive MFM frequency images and corresponding phase offset images with $V_{dc}$= -2 V. **c,d,** the corresponding phase error images. **e,** Averaged line scans (in the dashed box) perpendicular to domain patterns for two successive scans at $V_{dc}$= -2 V illustrate how the fine domain structure is altered by the MFM tip.

**Fig. 4 | Magneto-electric force microscopy (MeFM) experiments above the bare LAO near the top electrode. a**, Sketch of the experimental setup. A combined ac and dc bias is applied to the interface with the top gate grounded. The ac bias is at a fixed frequency equal to the cantilever's free resonant frequency. Unlike MFM, the tip is not driven mechanically. The tip is electrically isolated and kept at a constant height above the sample surface during imaging. **b,** Topographic image of the sample where MeFM images are acquired. Image size is 1.5 μm × 1.5 μm about ~5 μm away from the edge of the top electrode. **c,** MeFM amplitude images acquired with $V_{dc}$ increasing from -3.5 V to 0 V. Each image is shown with the average value subtracted: 0.26 mV, 0.19 mV, 0.14 mV, 0.09 mV, 0.06 mV, 0.07 mV respectively.

**Fig. 5 | Proposed magnetic phase diagram for LAO/STO**. Four distinct phases exist as a function of the total density of interface electrons. **a,** Low-density limit, e.g., below the critical thickness of LAO. **b,** Ferromagnetic (FM) phase with antiferromagnetic coupling between local moments and semi-extended states (both $d_{xy}$ orbitals). **c,** Non-magnetic conductive state. **d,** Second magnetic phase M* above the Lifshitz transition. **e,** Phase diagram versus interface electron density and temperature *T*.

Supplementary Information available online

Acknowledgments

We gratefully acknowledge helpful comments from and discussions with Cheng Cen, Guanglei Cheng, Sergey Frolov, Lior Klein, Roman Lutchyn, Chetan Nayak, Roger Proksch, Evgeny Tsymbal, and Eli Zeldov. This work was supported by ARO MURI W911NF-08-1-0317 (JL), AFOSR MURI FA9550-10-1-0524 (CBE, JL) and FA9550-12-1-0342 (CBE), and grants from the National Science Foundation DMR-1104191 (JL), DMR-1124131 (CBE, JL), and DMR-1234096 (CBE).

Author contributions

F.B. designed the devices, performed the experiments and wrote the manuscript. M.H. fabricated the devices. C.W.B and S.R. grew the samples. C.B.E. reviewed the manuscript. P.I. edited the manuscript. J.L. suggested and designed experiments and co-wrote the manuscript.

Author information

Correspondence and requests for materials should be addressed to J.L. (jlevy@pitt.edu)

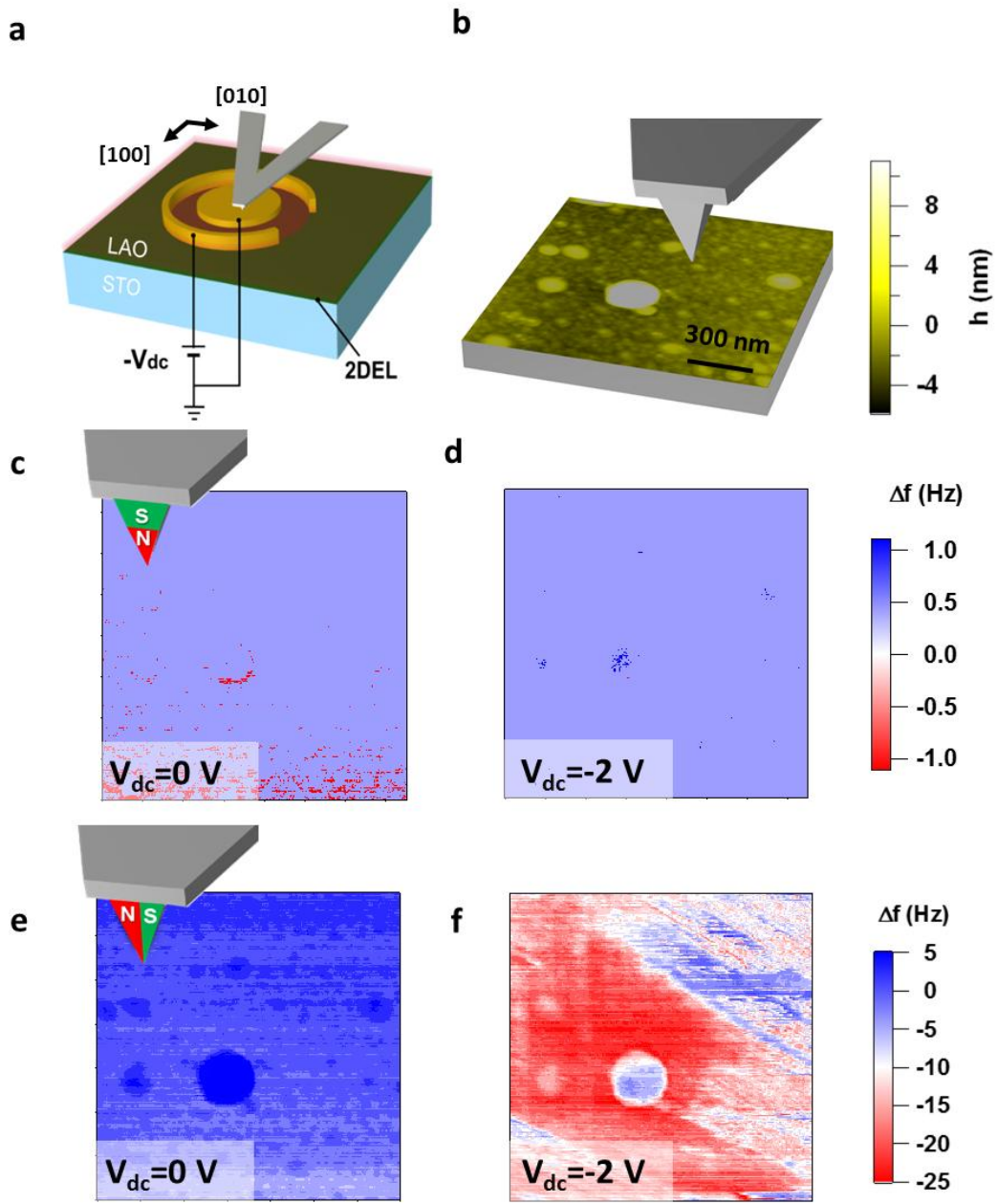

**Bi et al, Figure 1**

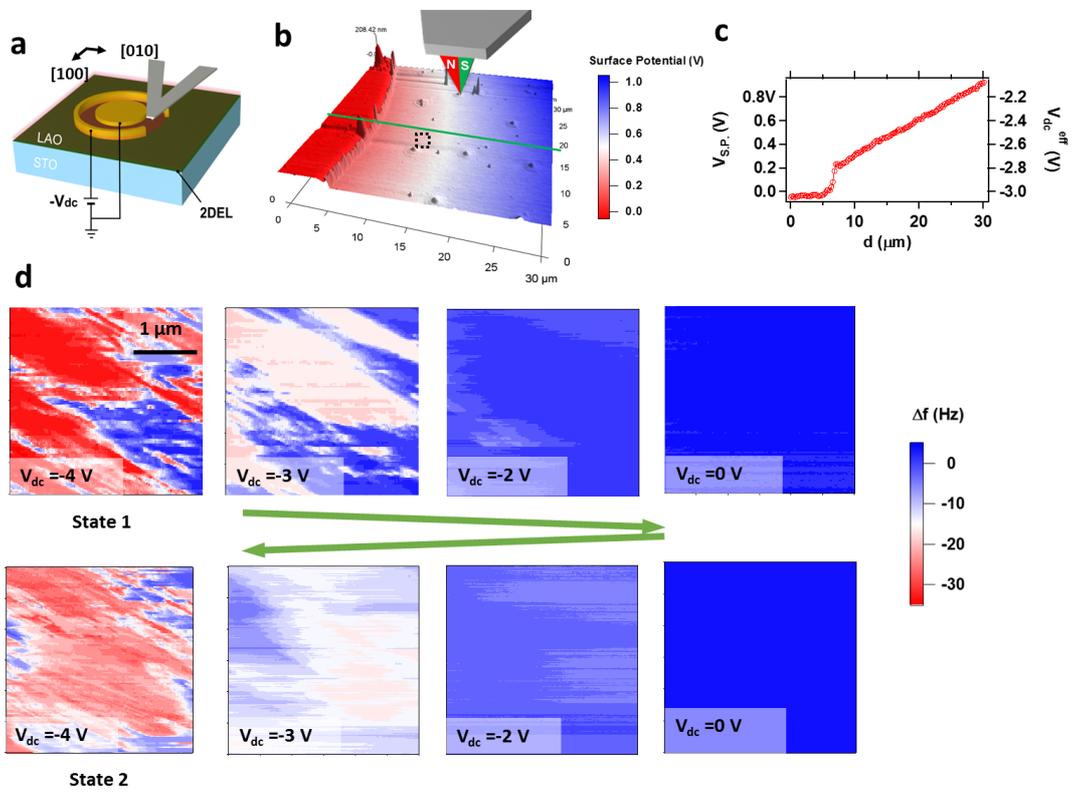

Bi et al, Fig. 2

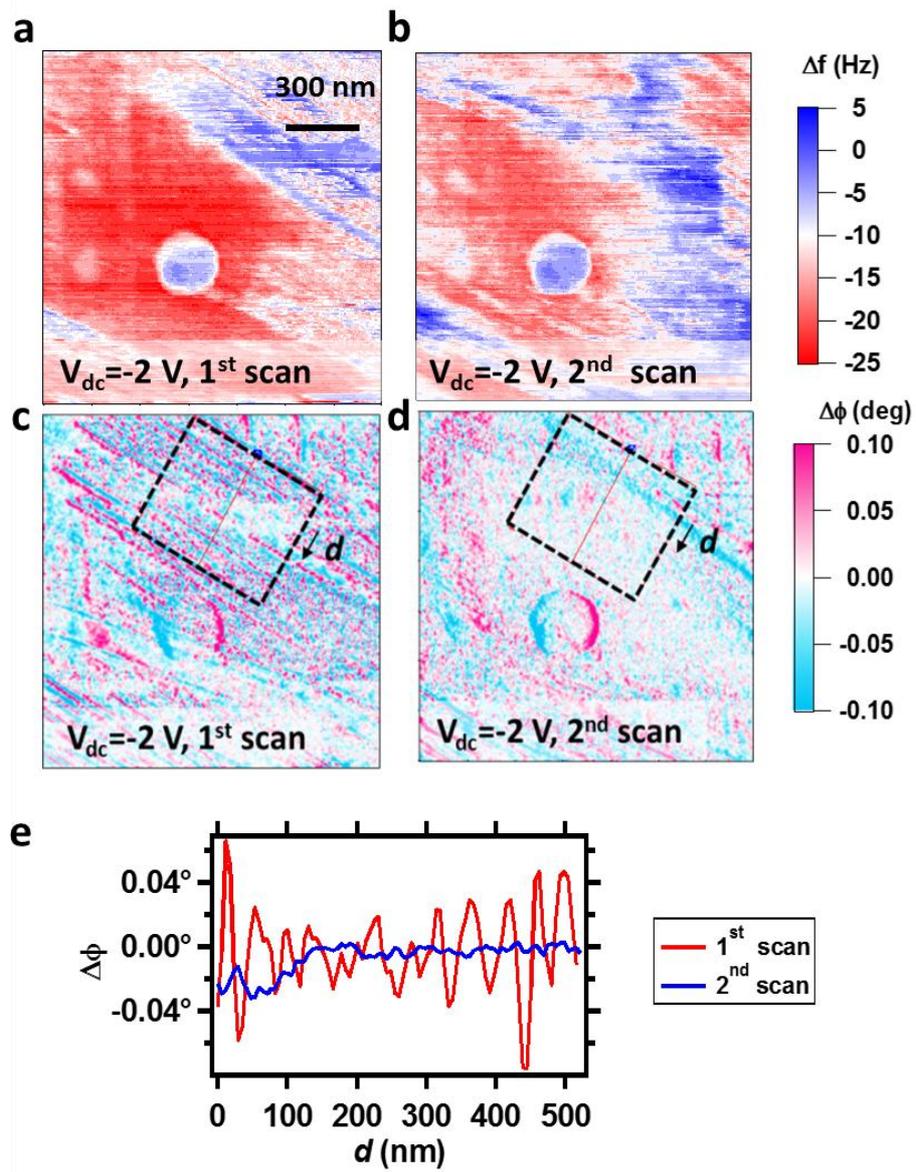

Bi et al, Fig. 3

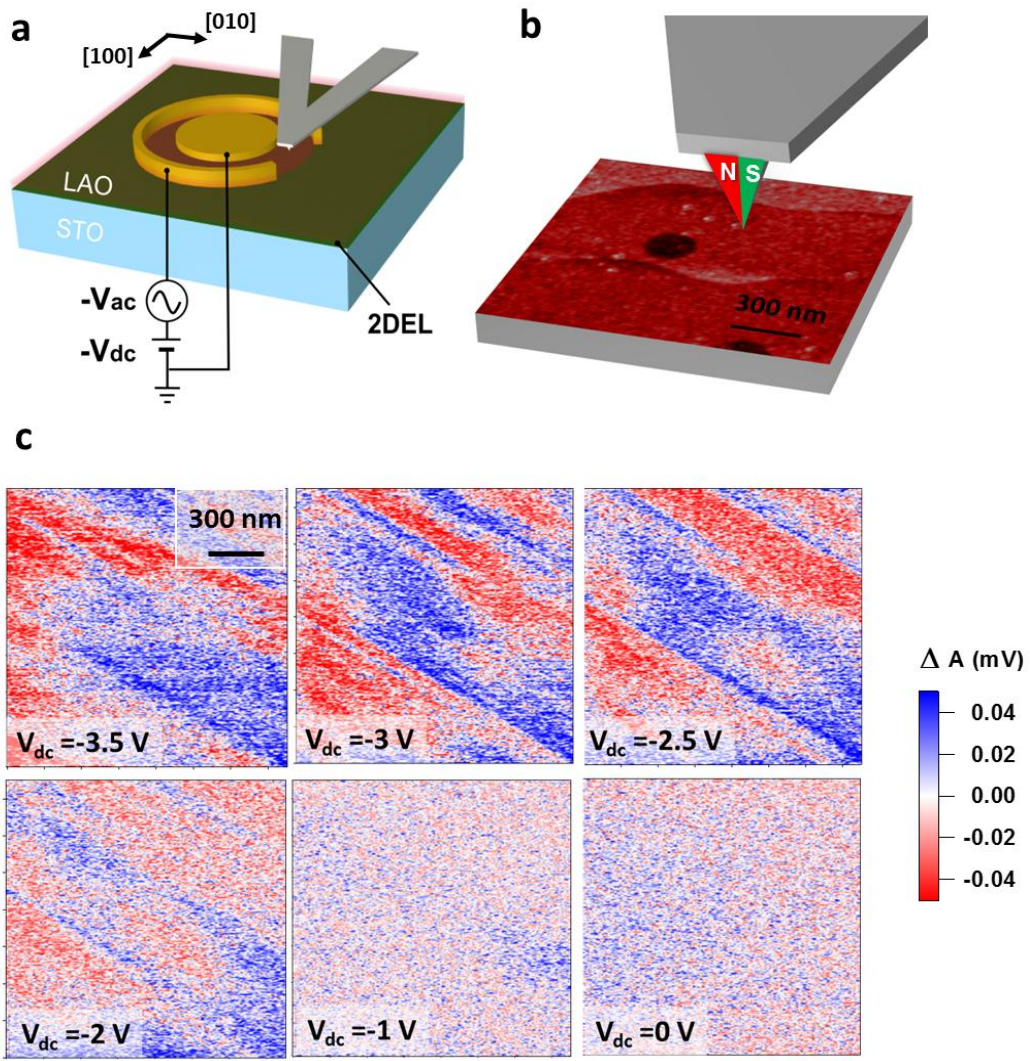

Bi et al, Fig. 4

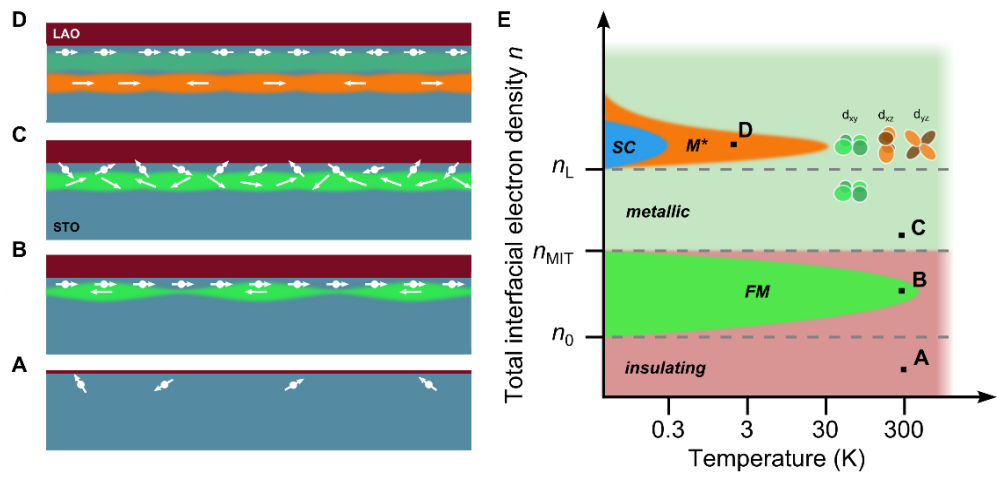

**Bi et al, Fig. 5**

SUPPLEMENTARY INFORMATION

**DEVICE FABRICATION**

The geometry of devices are discussed in the Methods section in the main text. The capacitor electrodes are deposited on the LAO/STO samples via DC sputtering. The arch-shaped electrodes are prepared by creating 25 nm trenches via Ar-ion milling, followed by deposition of 4 nm of Ti and 30 nm of Au. A series of metallic circular top gates (4 nm Ti and 40 nm Au) are deposited on the LAO surface. The arc-shaped electrodes have a width of 20 µm and fixed separation of 50 µm to the edge of the circular top gates. The entire sample (5 mm × 5 mm × 0.5 mm) is affixed to a ceramic chip carrier using silver paint. Electrical contacts to bonding pads on the device are made with an ultrasonic wire bonder using gold wires.

The investigated devices with parameters are summarized in Table S1.

Table S1: Sample information and device parameters.

|  | Top-gate diameter | LAO thickness | Growth conditions |
| --- | --- | --- | --- |
| Device A | 100 µm | 12 u.c. | Grow @ 780°C, $P(O_2)=10^{-5}$ mbar; Anneal @ 600 °C, $P(O_2)=$ 300 mbar |
| Device B | 200 µm | 12 u.c. | Grow @ 780°C, $P(O_2)=10^{-5}$ mbar; Anneal @ 600 °C, $P(O_2)=$ 300 mbar |
| Device C | 400 µm | 12 u.c. | Grow @ 550°C, $P(O_2)=10^{-3}$ mbar |

**DEVICES IMAGES AND SCANNING LOCATION LIST**

Fig. S1a shows the fabricated sample loaded in the chip carrier with electric connections bonded. In the red dash line enclosed region, a zoomed in optical image is given in Fig. S1b. A further zoom in pictures of Device A, B and C in Table. S1 are shown in Fig. S1c-e respectively with all the scan locations indicated by the markers. A table of figures and corresponding locations is provided as Fig. S1f.

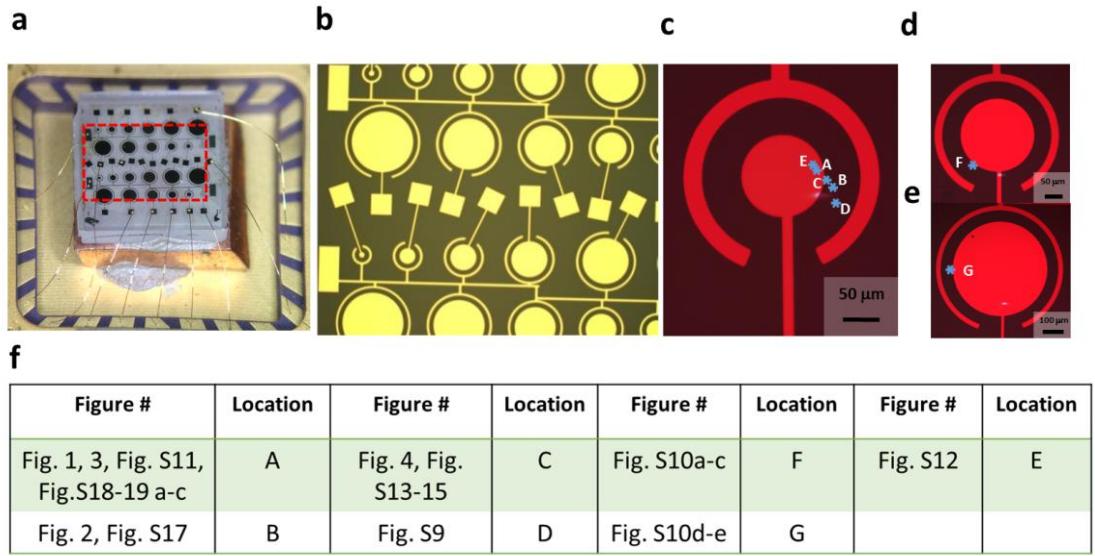

| Figure # | Location | Figure # | Location | Figure # | Location | Figure # | Location |
|---|---|---|---|---|---|---|---|
| Fig. 1, 3, Fig. S11, Fig.S18-19 a-c | A | Fig. 4, Fig. S13-15 | C | Fig. S10a-c | F | Fig. S12 | E |
| Fig. 2, Fig. S17 | B | Fig. S9 | D | Fig. S10d-e | G | | |

Fig. S1. **Devices images and scanning location list**. **a-b,** Pictures of sample and devices under optical microscopy. **c-e,** images of Device A-C. **f,** route map of figures and investigated locations.

## CV CHARACTERIZATION OF DEVICE

CV measurements are performed using a commercial capacitance bridge (Andeen-Hagerling 2500A) between the top gate and electrode contacting the interface on Device A. The ac excitation voltage is given by $V_C$=10 mV, $F_C$=1 kHz and $V_{dc}$ sweep at a rate $dV_{dc}/dt$=0.067 V/s. As shown in Fig. S2, the capacitance becomes suppressed as $V_{dc}$ is reduced from 1 V to -4 V, indicating a MIT as the gate bias is varied. The transition takes place within the range 0 V to -2 V and is hysteretic.

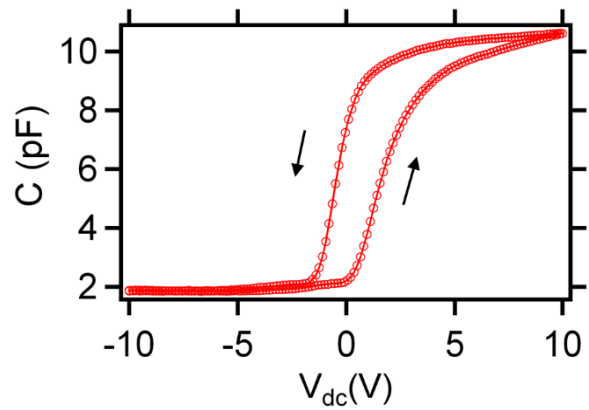

Fig. S2. **CV characterization.** Measurement is performed on Device A with ac excitation given by $V_{ac}$=10 mV and $F_c$=1 kHz.

**LEAKAGE CURRENT**

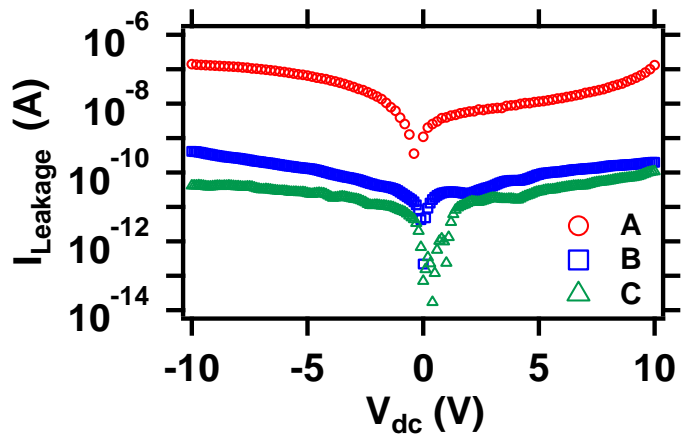

Fig. S3. **Leakage current between top electrode and interface.** Measurements are performed for Devices A-C.

**CANTILEVER INFORMATION AND CALIBRATION**

The magnetic tip (Manufacturer: Asylum Research, Model: ASYMFM) is coated with CoCr. The magnetic coating on the tip has a coercive field of 400 Oe and a total magnetization $10^{-13}$ emu. The mechanical properties of the cantilever, such as spring constant and resonant frequency, are obtained by performing thermal spectroscopy. During this measurement, the AFM monitors the cantilever's fluctuation as a function of time. Based on the time-domain measurement, the AFM extracts the power spectral density (PSD), which is shown in Fig. S4. The cantilever's thermal noise response with respect to frequency is then fitted to a simple harmonic oscillator equation, from which we obtain the spring constant $k = 1.2$ N/m and quality factor $Q = 136$.

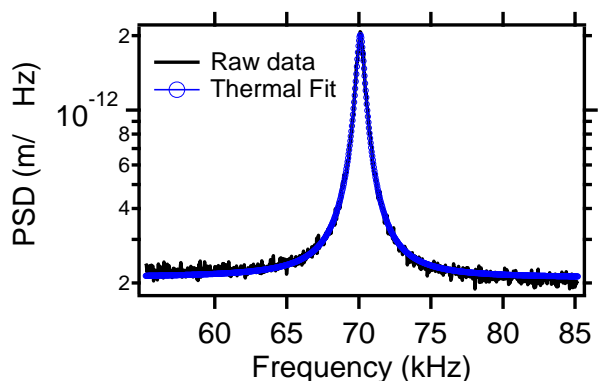

Fig. S4. **Thermal spectroscopy of the MFM cantilever.**

**TIP MAGNETIZATION AND VERIFICATION**

The MFM tip is magnetized using an electromagnet in Fig. S5, which can apply a uniform magnetic field up to 2000 Oe. During the magnetization, tip is placed in the gel box at the middle of two electromagnets. By rotating the tip holder, we can magnetize the tip horizontally or vertically.

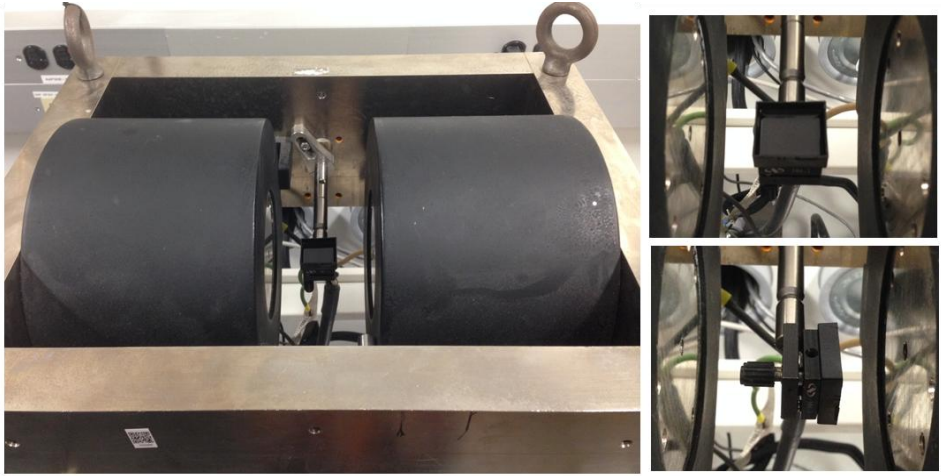

Fig. S5. **Electromagnetic system used to magnetize the MFM tip**

To verify that the tip is properly magnetized, MFM measurements are performed on a commercially available (Bruker, model "MFMSAMPLE") test sample with periodic in-plane magnetic domain structures and out-of-plane domain walls. During the MFM imaging, the tip-surface distance Δh is kept at 50 nm. With the tip vertically magnetized, MFM images are acquired with the sample rotated clockwise 0° and 37° (Fig. S6a-b). Since the tip magnetization is perpendicular to the sample, Fig. S6a-b should show the same contrast, independent of the angle. A quantitative analysis of the averaged section profiles perpendicular to the domain wall is demonstrated in Fig. S6c show good agreement with the expected result.

The horizontal tip magnetization is checked using a similar method. Based on Eq.1c, $\Delta f \propto m \cdot \frac{\partial B}{\partial z} \cdot \cos\theta$, where $\theta$ is the angle between tip and sample's magnetization orientation. Changing relative angle between sample and tip will change the magnetic signal. Fig. S6d-e are MFM images with sample rotated clockwise 0° and 48°. Fig. S6e shows a significantly lower signal than Fig. S6d, which is also directly displayed in the Fig. S6e. The signal contrast ΔF can be simply defined as the averaged difference between peaks and vales of the curves in Fig. S6e, from which $\Delta F_{0°}$= 16.8 Hz and $\Delta F_{48°}$= 7.3 Hz. Therefore $\Delta F_{0°} / \Delta F_{48°} = 2.3$. On the other hand, $\Delta F_{0°} / \Delta F_{48°} = \cos\theta_0 / \cos\theta_{48}$, where $\theta_0$ and $\theta_{48}$ are the angle between tip and sample magnetization in Fig. S6d and Fig. S6e. With $\theta_0=15°$ and $\theta_{48}=63°$, $\Delta F_{0°} / \Delta F_{48°} = \cos 15° / \cos 63° = 2.1$, which is close to the result (2.3) from Fig. S6e. Therefore, Fig. S6d-f confirms that the tip is properly magnetized in plane.

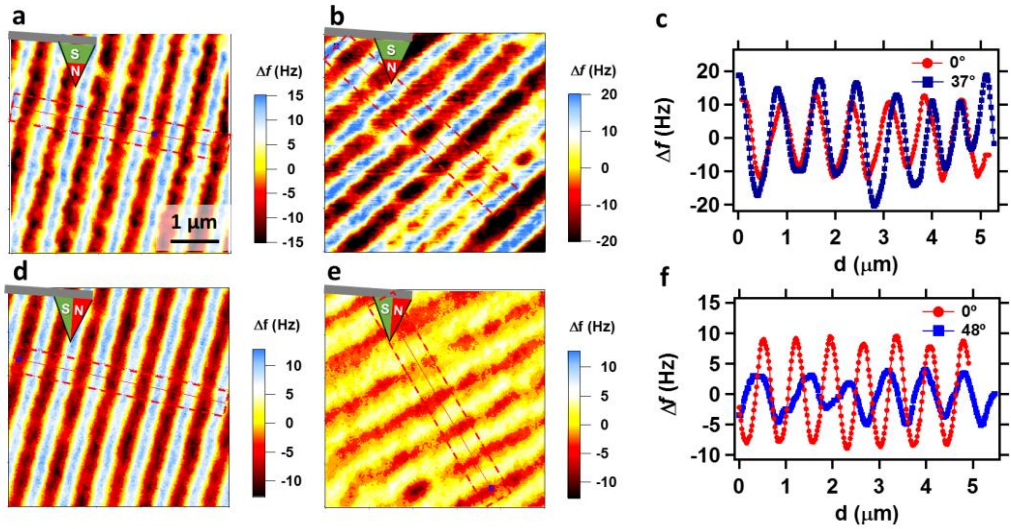

Fig. S6. **MFM images on reference sample. a-b**, MFM images with sample rotated clockwise at 0° and 37° respectively. Tip is magnetized vertically as it is indicated in the figure. **c**, Averaged section profile along the red line in **a** and **b**. **d-e,** MFM images obtained with sample rotated clockwise at 0° and 48° respectively. Tip is magnetized horizontally. **f,** Averaged section profile along the red line in **d** and **e**.

### MAGNETIC FIELD ENVIRONMENT

Fig. S7 illustrates the scanning environment in a commercial AFM. There is a magnetic holder (blue) directly attached to the scanner. The sample is loaded in a chip carrier on a printed circuit board (grey). A circular steel specimen disc is glued to the bottom of PC board (dark grey). The magnetic force between the steel disc and magnetic holder ensures that the sample sticks to the scanner during the scanning.

Using a magnetic sensor, the measured magnetic field on the sample surface is: $B_z = 1.1\,\text{Gauss}$, $B_x = 1.8\,\text{Gauss}$, $B_y = 0.8\,\text{Gauss}$. This static magnetic field (which includes the Earth's contribution) exists during all of the experiments described. For Fig. S9, the sample rotates with the steel disc, while the magnetic holder remains fixed in place.

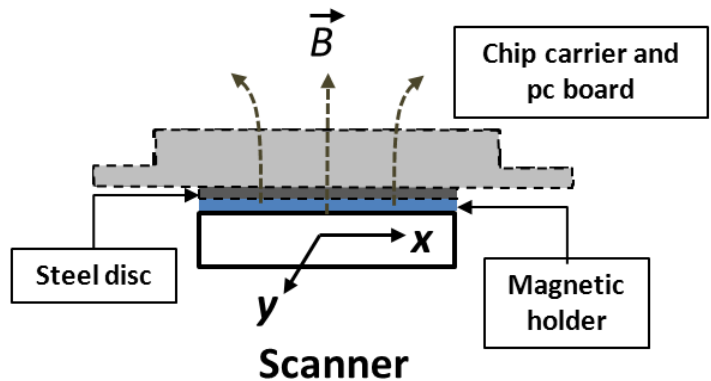

Fig. S7. **Schematic showing the AFM scanning environment.** There is a weak (~1 Oe) magnetic field from the magnetic holder.

### KFM SURFACE POTENTIAL MEASUREMENT

Kelvin probe force microscopy is employed to measure the surface potential across the top electrode and bare LAO. The electric drive amplitude is 100 mV at 78.8 kHz. The tip-surface distance is kept at 20 nm and the top electrode is grounded during the KFM imaging.

In the first measurement, the electrode is grounded to the interface ($V_{KFM} = 0$). KFM is employed to measure the work function difference between top electrode and LAO (Fig. S8b). Then the dc bias is increased to $V_{KFM} = 3$ V and a subsequent KFM image is obtained under a same scan conditions (Fig. S8c). The surface potential distribution shown in Fig. 2 is obtained by subtracting the $V_{KFM}=0$ KFM image from the $V_{KFM}= 3$ V image.

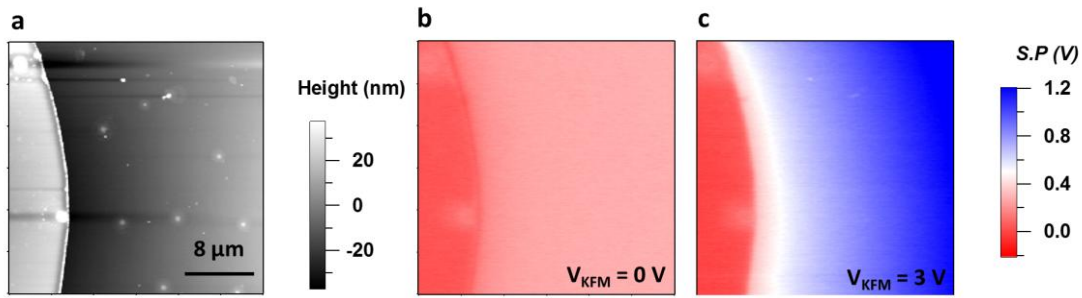

Fig. S8. **KFM characterization. a,** Height image over a 30 μm × 30 μm area across the boundary of top electrode. **b-c,** KFM images with $V_{KFM} = 0$ V and 3 V respectively, where $V_{KFM}$ is the dc bias that applied to the electrodes contacting the interface.

**MFM CONTRAST VERSUS IN-PLANE TIP MAGNETIZATION DIRECTION**

Changing the relative angle between the tip and the sample can help to identify the orientation of the magnetic domains. A series of MFM images is shown in Fig. S9 at $V_{dc}=-1$ V, where the relative angle $\theta$ between the MFM tip and the sample is varied. The tip magnetization is fixed along the $y$ direction and the sample is rotated clockwise with various angles ranging from 0° to 42°. The MFM amplitude contrast is largest at $\theta=0°$. As the sample rotates, the stripe-shaped feature rotates by approximately the same angle; however, the magnitude of the signal decreases. At $\theta = 42°$, the stripe-shaped features vanish, resulting in a spatially uniform signal over the measured area.

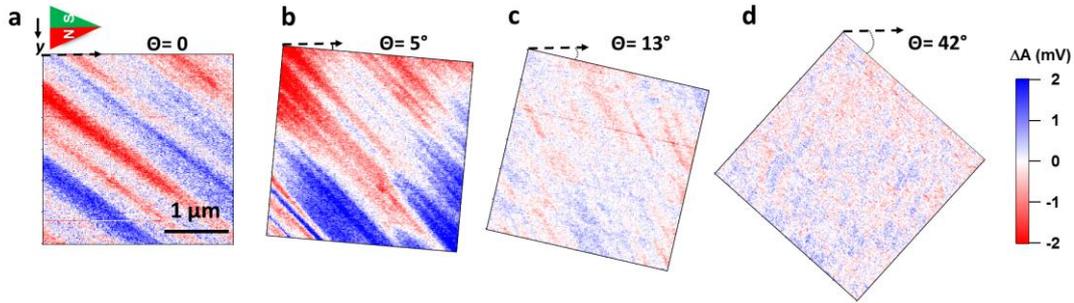

Fig. S9. **MFM images acquired for various relative angles between the tip magnetization and the sample orientation.** The slope detection method is applied with the cantilever driven at a fixed frequency 67 kHz slightly off resonance. The tip magnetization direction is kept unchanged while the sample is rotated clockwise with angle $\theta = 0°$ **a**, 5° **b**, 13° **c**, 42° **d**.

**MFM ON DEVICE B AND DEVICE C**

MFM experiments performed on Devices B and C yield similar results. The separation between tip and surface is 50 nm. Fig. S10 shows MFM images over a 5 μm × 5 μm area on a bare LAO surface close to the boundary of the top electrodes of the two devices.  Fig. S10a-c shows the data from Device B while Fig. S10d-f are the results from Device C. The topographic images for the investigated area on Device B and C are demonstrated in Fig. S10a and Fig. S10d, respectively.  Both MFM images Fig. S10b and Fig. S10e show the same characteristic stripe domains under low carrier density, similar to results observed in Device A in the main text. At the same time, Fig. S10c and Fig. S10f show significantly diminished contrast for $V_{dc}=1$ V bias.  This results agree qualitatively with the observations for Device A.

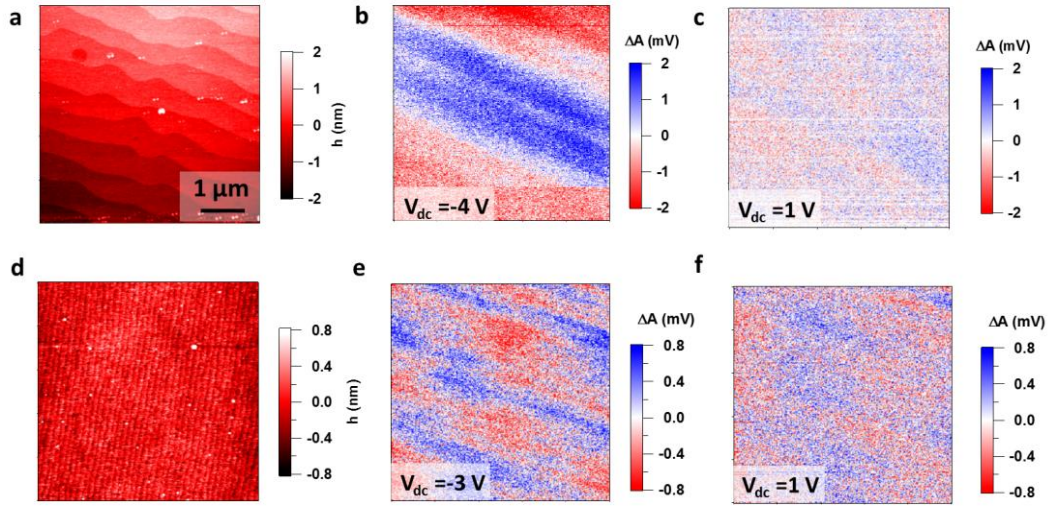

Fig. S10. **MFM on bare LAO surface close to the top electrode for Device B and C.** The image size is 5 μm. **a** and **d** shows surface morphology. **b-c** and **e-f** are MFM amplitude images under two different dc bias for Device B and C. **b** and **c** are offset 0.479 V and 0.403 V. **e** and **f** are offset 0.466 V and 0.416 V.

**MeFM EXPERIMENTS OVER TOP ELECTRODE**

MeFM experiments over the exposed LAO layer are described in the main text. Here, MeFM experiments are presented for an area over the top electrode. The experiment set up is depicted in Fig. S11a. The combined ac and dc voltage is applied to the interface with $V_{ac} = 0.14$ V at $f = 70.6$ kHz and $V_{dc}$ held at various fixed values. The separation between tip and top electrode is 20 nm. Fig. S11a illustrated the MeFM scanning above top electrode on device A. Fig. S11b shows the surface height image. The MeFM amplitude and phase images are included in Fig. S11c. The same measurements are also performed over another location on the same device (Fig. S12).

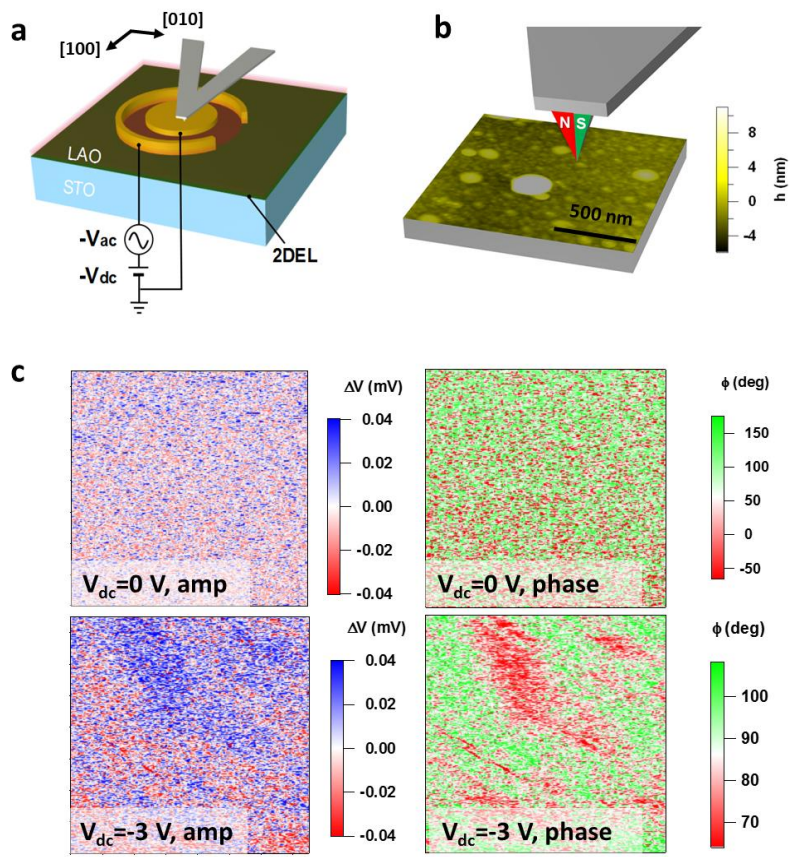

Fig. S11. **MeFM scanning above top electrode over 1.4×1.4 µm² square on Device A. a,** Sketch of experiment setup. **b,** The surface height image. **c,** MeFM amplitude and phase images with $V_{dc}$ =0 V to -3V.

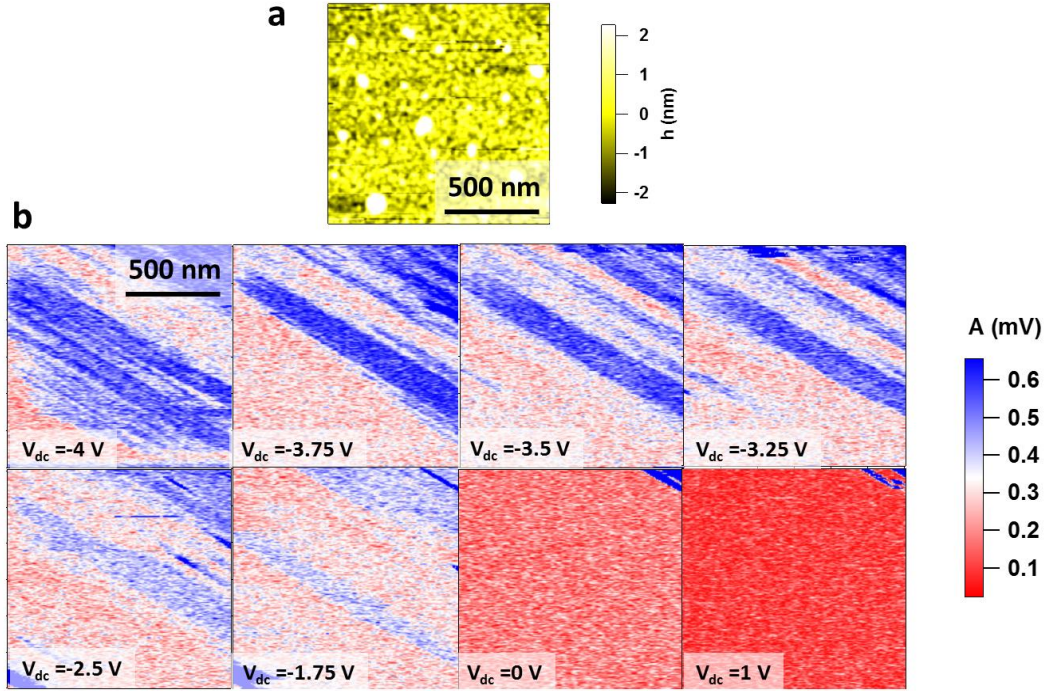

Fig. S12. **Another MeFM scanning over 1.2×1.2 µm² square on Device A. a,** the surface height image. **b,** a sequence of MeFM amplitude images as $V_{dc}$ increases from -4 V to 1V.

## MEFM WITH $V_{dc}$ SWEEPS IN OPPOSITE DIRECTIONS

MeFM experiments are also performed with $V_{dc}$ cycling from -0.5 V to -3.5 V and then back to -0.5 V. Measurements are done over the same area. The experiment setup is shown in Fig. 4a. A bias $V_{IF} = -(V_{dc} + V_{ac}\sin(2\pi ft))$ is applied to the interface with $V_{ac} = 0.14$ V and $f = 70.6$ kHz. The tip remains at a constant height 20 nm above the LAO surface and scanned over a 1.5 µm square area on Device A. Fig. S13a-f show MeFM amplitude images as $V_{dc}$ sweeps from -0.5 V to -3.5 V. The corresponding MeFM phase images are shown in Fig. S14a-f.

After a two-hour pause, MeFM images are obtained with $V_{dc}$ increasing from -3.5 V to -0.5 V (Fig. S13-14g-l). The results confirm that magnetic domain features are only observed when the mobile carrier density at the interface is low. Fig. S13-14 shows hysteretic behavior during the $V_{dc}$ cycling, which is also observed in CV measurements.

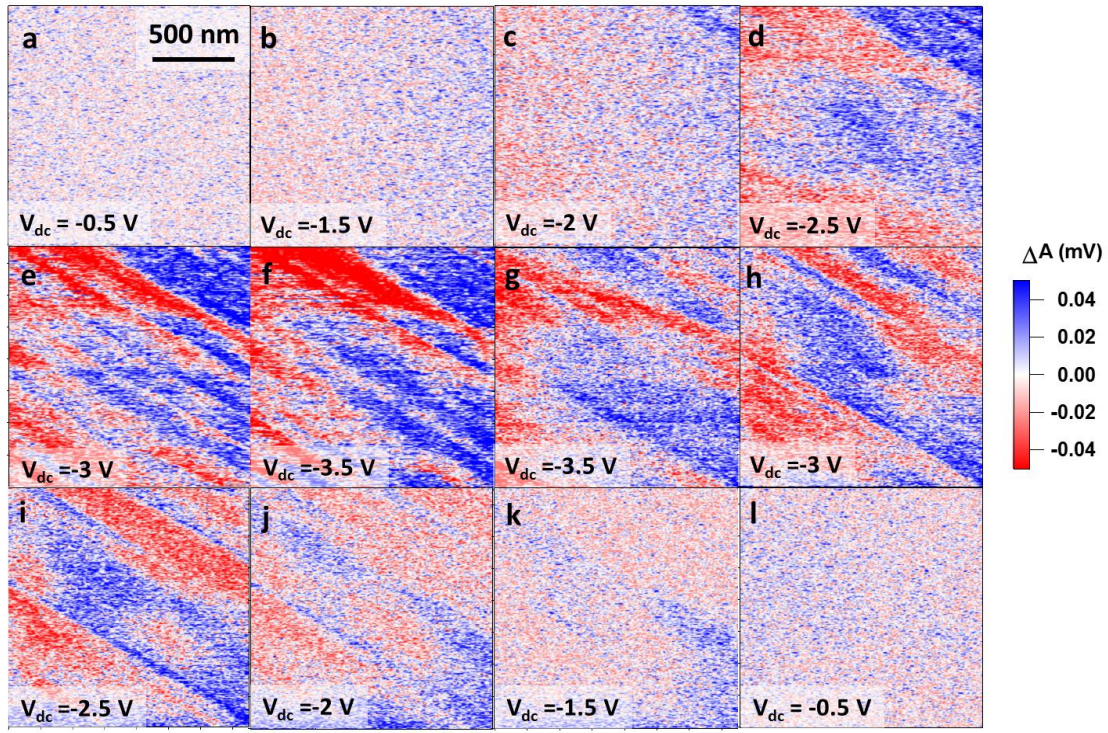

Fig. S13. **MeFM (amplitude) scanning over a 1.5 µm×1.5 µm area on LAO close to top electrode.** $V_{dc}$ sweeps from -0.5 V to -3.5 V. **a-f**, and after about 2 hours, $V_{dc}$ sweeps back from -3.5 V to -0.5 V **g-l**. In order to show the contrast of all images in the same color scale range, the amplitude of each image **a-l** offsets 0.04 mV, 0.05 mV, 0.09 mV, 0.2 mV, 0.3 mV, 0.4 mV, 0.3 mV, 0.2 mV, 0.1 mV, 0.09 mV, 0.06 mV, 0.06 mV. This indicates the magnetoelectric susceptibility increases as $V_{dc}$ decreases. Results are shown for Device A.

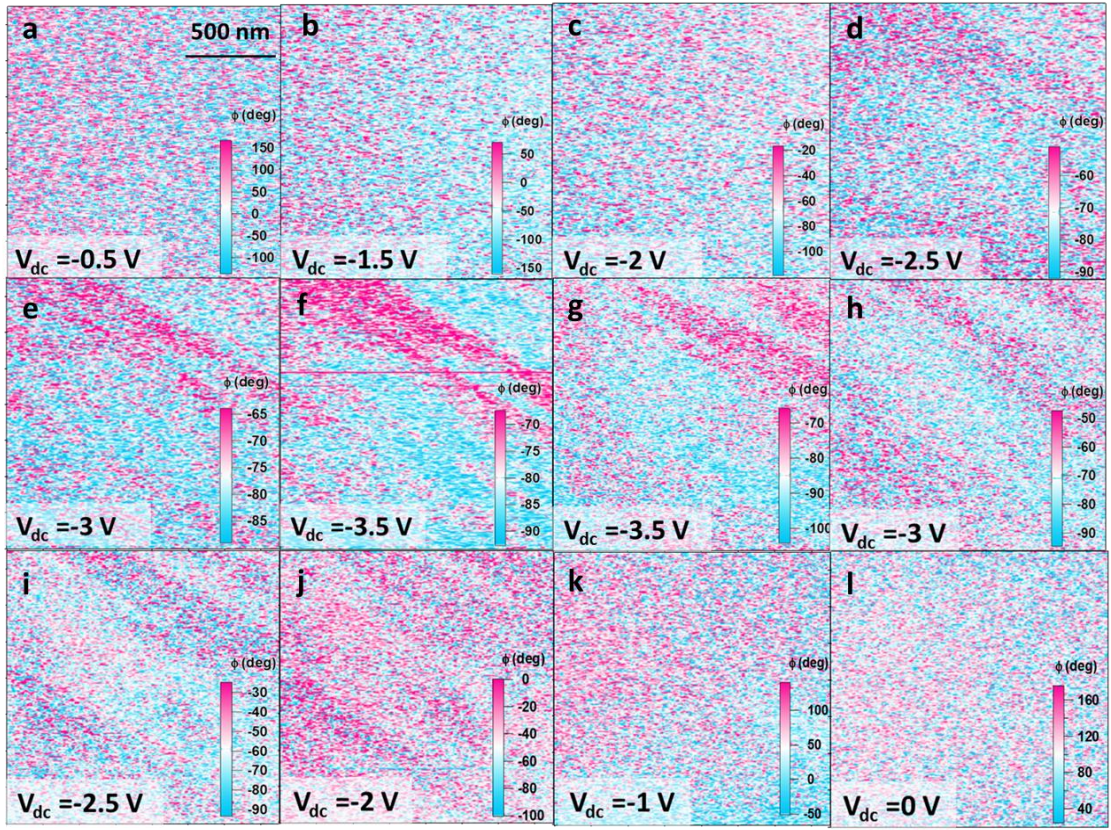

Fig. S14 **P****hase channel of MeFM experiments** (Fig. S13) above the bare LAO near the top electrode.

**2D FOURIER ANALYSIS**

In order to analyze the sharpness of the domain features, 2D Fourier analysis is applied to a series of MeFM images taken under varied $V_{dc}$ from -3.5 V to 0 V. The corresponding 2D fast Fourier transform (FFT) spectra are shown in Fig. S15b. There is a sharp peak in the 2D FFT image for $V_{dc} = -3.5\,\text{V}$. As $V_{dc}$ increases, the peak is gradually suppressed, vanishing for $V_{dc} \geq -1\,\text{V}$. To quantitatively analyze the peak in 2D FFT, a 1D section is plotted along the peak for each dc bias (Fig. S15c). The FFT results for $V_{dc} \geq -0.5\,\text{V}$ are used to normalize these profiles. The curves in Fig. S15c are intentionally offset by one decade each for clarity. In Fig. S15c, the black dashed line represents the full width at half maximum (FWHM) for each curve and the data points are defined as cut off $k_0$. As $V_{dc}$ increases, $k_0$ decreases, which suggests the corresponding length scale $L_0 \sim 1/k_0$ increases. The $L_0$ is associated with the contrast sharpness in Fig. S15a, which quantitatively shows the domain wall width.

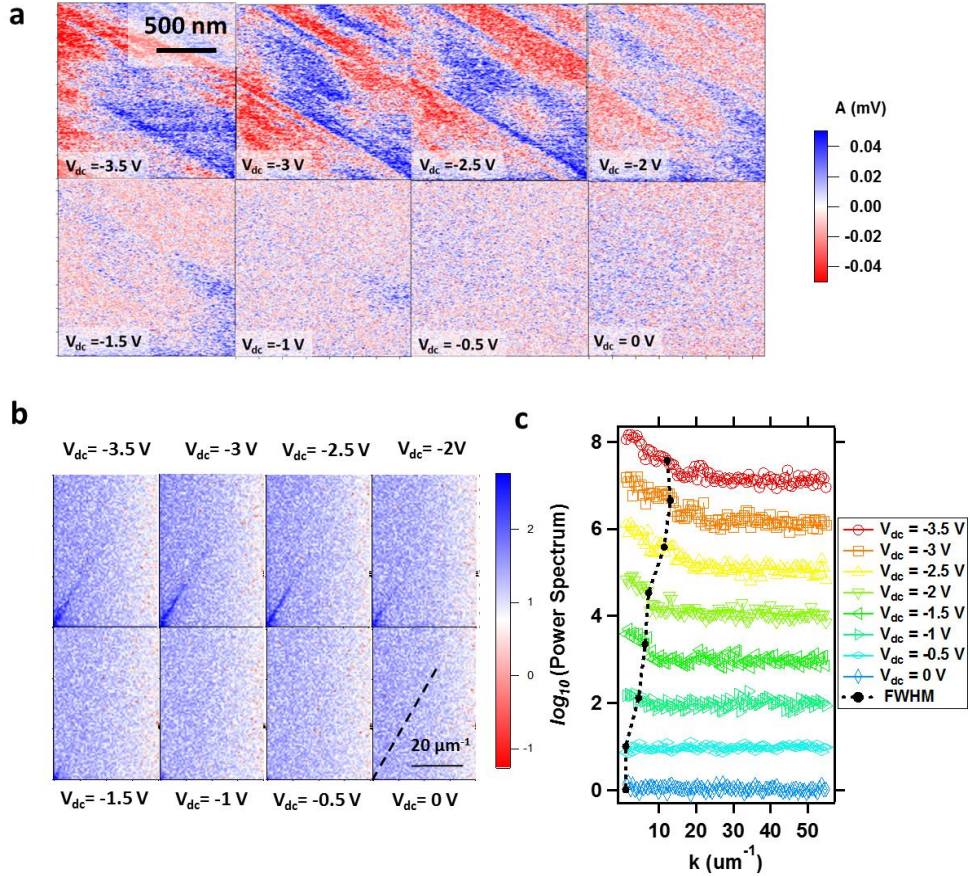

Fig. S15. **2D Fourier analysis of MeFM results. a,** MeFM images taken under different $V_{dc}$. The amplitude of each image is offset 0.26 mV, 0.19 mV, 0.14 mV, 0.09 mV, 0.06 mV, 0.06 mV, 0.06 mV, 0.07 mV respectively. The scan size is 1.5 µm × 1.5 µm. **b,** 2D FFT results for each image in **a**. For each FFT image, the *x* axis has range from 0 to 40 µm$^{-1}$ and the *y* axis has range from 0 to 60 µm$^{-1}$. **c,** FFT section profile along the dash line in **b**. We have averaged the FFT results for $V_{dc} \geq -0.5\,\text{V}$ as the background. Each power spectrum is normalized by the value at $V_{dc} = 0\,\text{V}$. The offset between curves is 10 dB. The black dotted line in **c** is a guide to the eye, and indicates the full width half maximum (FWHM) of the FFT curve.

**HISTOGRAM ANALYSIS OF MFM IMAGES ON MAGNETIC DOMAIN BOUNDARY**

The dependence of the magnetic contrast on gate voltage $V_{dc}$ can be quantified by performing a histogram analysis on a sequence of images (Fig. S16). The average value of the two central domains can be identified by the peaks in the double-valued histrogram. Fig. S16c shows the dependence on gate voltage, which is approximately linear in the range $-3.5\text{ V} \leq V_{dc} \leq -1.5\text{ V}$. Scans at increasingly negative gate voltages are not performed due to a rapid increase in leakage current through the LAO.

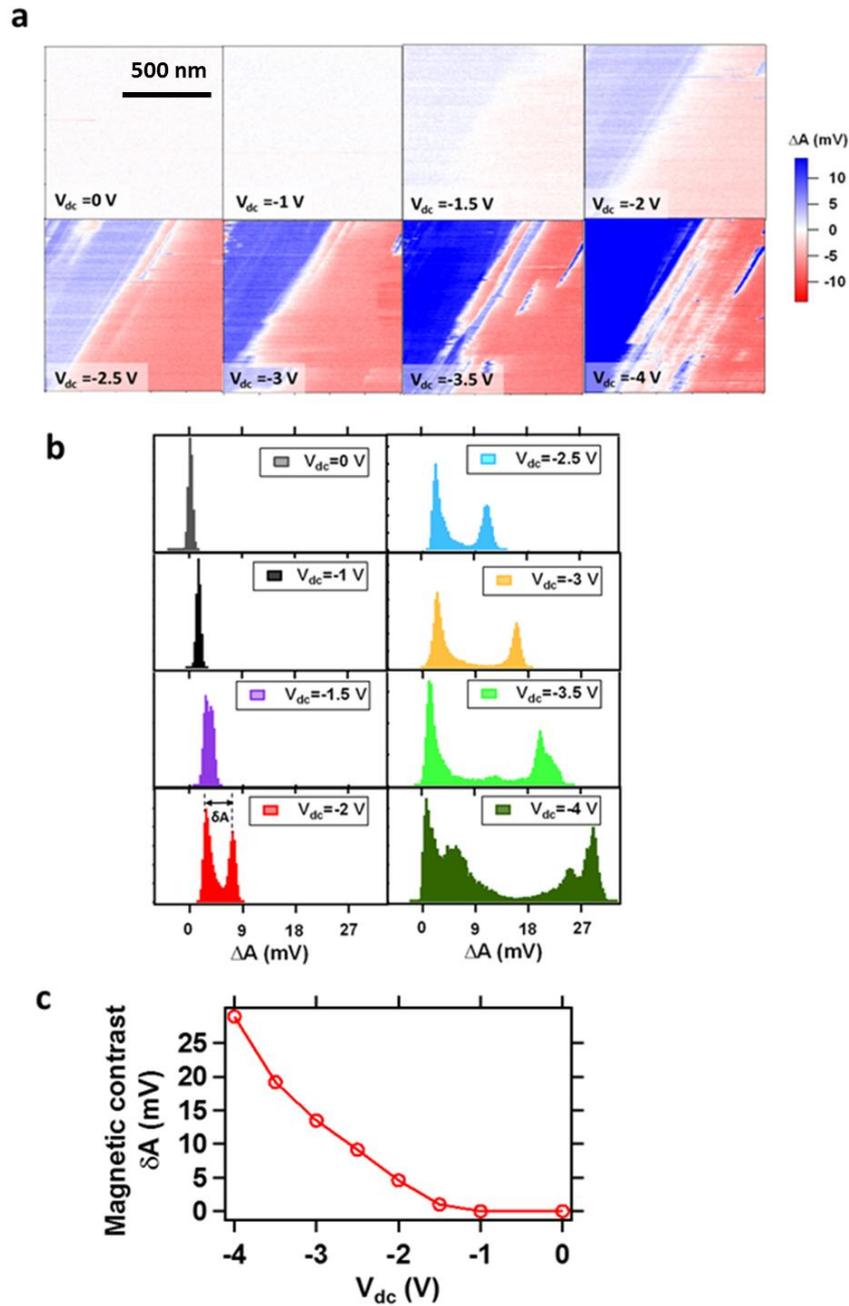

Fig. S16. **Histogram analysis of MFM images on magnetic domain boundary.** **a,** MFM images on LAO near top electrode. (1 μm×1 μm). For $V_{dc}$ from 0 V to -4 V, each image is offset 565 mV, 567 mV, 568 mV, 569 mV, 569 mV, 573 mV, 573 mV, 573 mV respectively. **b,** histogram analysis of MFM images in **a**. The peak in histogram of $V_{dc}$=0 corresponds to ΔA=0. The observed double peaks in histogram figures correspond to the different contrast of two magnetic domains in **a**. We therefore define the magnetic contrast to be the gap δA between the two peaks, which is plotted in **c** as a function of $V_{dc}$.

### AVERAGE MFM FREQUENCY SHIFT VERSUS $V_{DC}$

A sequence of MFM images is obtained with $V_{dc}$ sweeps from 4 V to -4V and then increases back to 4 V (Fig. S17a). For each dc bias, two successive MFM images are recorded. MFM image is obtained using frequency modulation method on exposed LAO near the top electrode on Device A. The scan size is 5 μm × 5 μm. In Fig. S17b, we calculated the average over each MFM image in Fig. S17a as a function of dc bias. The black arrow shows dc bias sweep direction. As $V_{dc}$ decrease below -2 V, the averaged frequency signal decreases rapidly, showing an increasing net magnetization within sample. Fig. S17b also shows a constant frequency drift, which can be observed from data change at the same dc bias. Such frequency drift could be due to temperature change in AFM chamber or something else. To get rid of the frequency drift, Fig. S17c plots the frequency change of two successive scanning at each dc bias. We treat the frequency drift rate to be the average of Fig. S17c, which is 1.0 Hz/image. The corrected $f_{ave}$ will be: $cor.\ f_{ave} = f_{ave} - 1 \times N$, where N is the image index number. The corrected figure is shown in Fig. S17d. The slight difference between $V_{dc}$ sweeps in two different directions is associated with charge hysteresis in the CV spectrum.

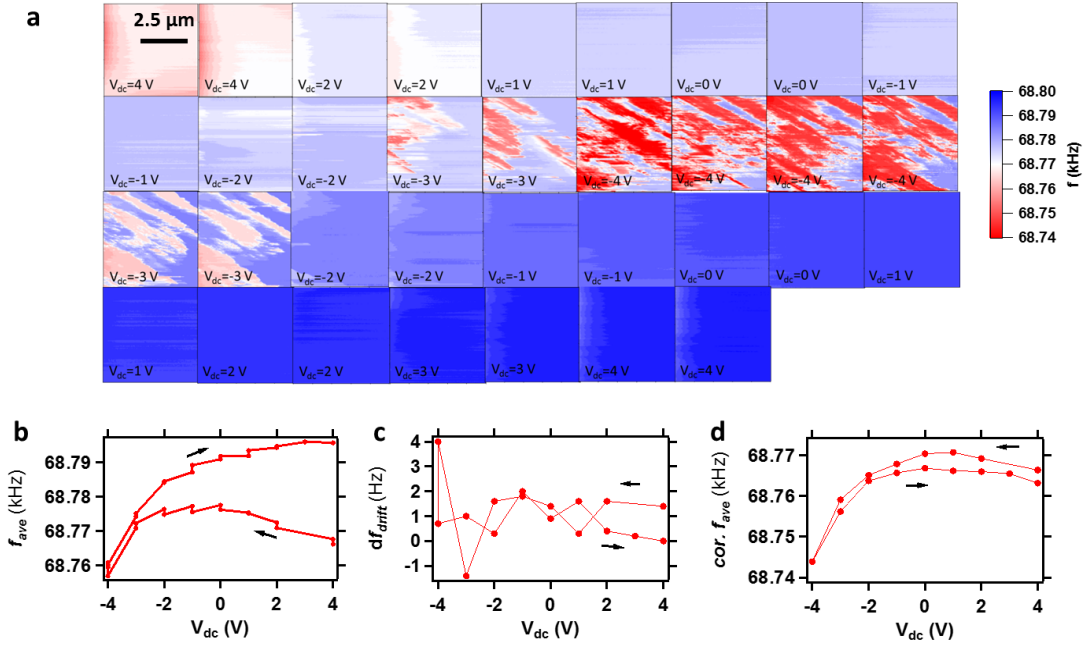

Fig. S17. **Average frequency analysis of MFM images with varied $V_{dc}$. a,** A sequence of MFM images is obtained with $V_{dc}$ sweeps from 4 V to -4V and then increases back to 4 V. The scan size is 5 μm × 5 μm. At each dc bias, successive MFM images are recorded. The MFM experiments are performed on exposed LAO near the top electrode on Device A. **b,** The frequency average of each image in **a** with respect to $V_{dc}$. **c** Frequency drift at each dc bias, obtained from the frequency change of two successive MFM images. The frequency drift rate is 1 Hz/image (22 min). **d,** Corrected figure **b** by subtract the frequency drift. The black arrow indicates the dc bias sweep direction.

## MFM EXPERIMENTS WITH A NON-MAGNETIC TIP

MFM and MeFM imaging are performed with a non-magnetic tip (heavy doped Silicon tip, Manufacturer: Aspire. Model: CFMR), shown in Fig. S18 and Fig. S19. In Fig. S18, the tip is driven at its resonant frequency $\omega_0/2\pi=70$ kHz at a height $\Delta h=20$ nm above the top electrode surface. In Fig. S19, the tip is also driven at the fixed resonant frequency 70 kHz. The tip-surface separation during scanning is $\Delta h = 20\,\text{nm}$ for Fig.

S19a-c and $\Delta h = 30\,\mathrm{nm}$ for Fig. S19d-f. With the non-magnetic tip, neither Fig. S18 nor Fig. S19 shows any magnetic signal or contrast in MFM and MeFM images when the interface is at low carrier density.

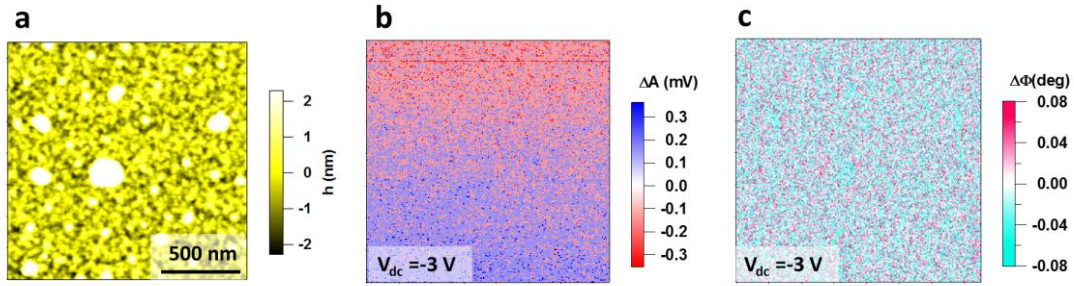

Fig. S18. **MFM experiments using non-magnetic tip on top electrode over a 1.5 μm×1.5 μm area. a,** Height image. **b-c**, MFM amplitude and phase image with $V_{dc} = -3\,\mathrm{V}$. The amplitude image is offset 0.357 V and phase image is offset 90°. Results are shown for Device A.

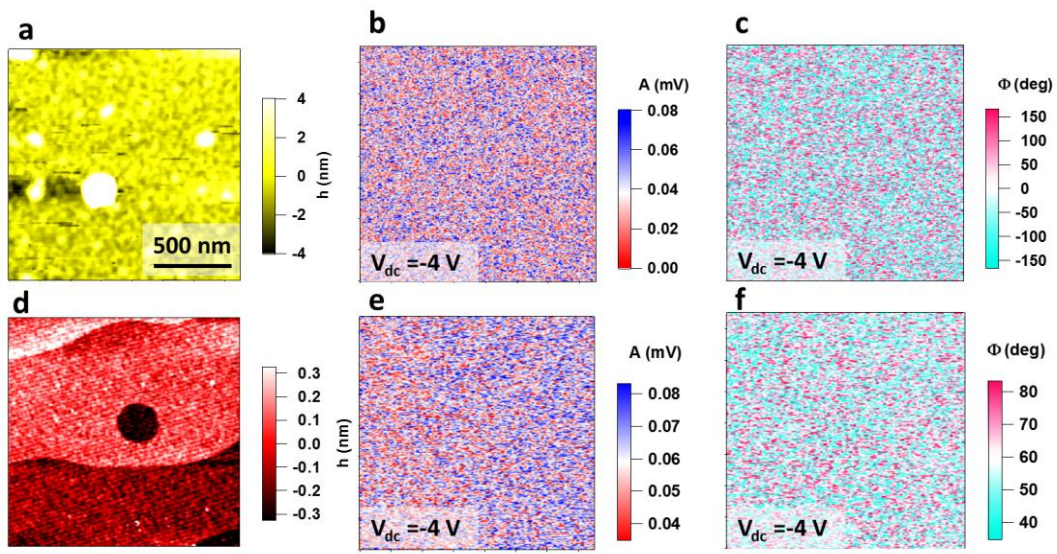

Fig. S19. **MeFM experiments on Device A using non-magnetic tip** over the top electrode **a-c** and over the LAO the boundary of the top electrode **d-f**. Scans

are taken over a 1.5 μm×1.5 μm area. Left images show topography while center and right images show corresponding MeFM signals (indistinguishable from noise).

## MFM EXPERIMENTS ON BARE STO SURFACE

Control experiment is performed on bare STO with horizontally magnetized MFM tip. The scan size is 3 μm. During the experiment, the tip is 40 nm above the surface. The MFM images in Fig. S20b and Fig. S20c do not show any magnetic signal, indicating the observed magnetic contrast in LAO/STO does not originate from the STO substrates.

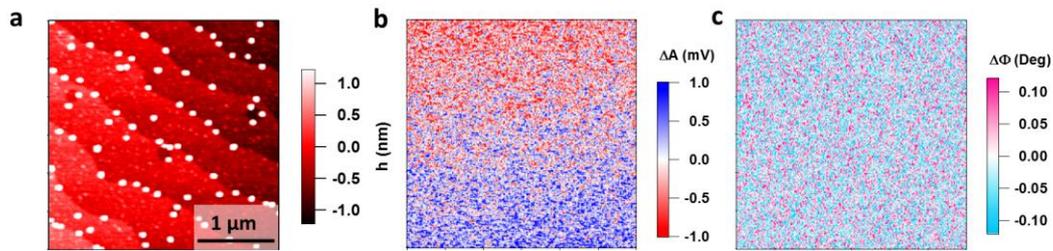

Fig. S20. **MFM experiment on bare STO with horizontally magnetized tip. a,** shows the height image over a 3 μm × 3 μm area. **b** and **c** show the MFM amplitude and phase image respectively.